\documentclass[%
 reprint,
 superscriptaddress,
 nofootinbib,
 twocolumn, 
 amsmath,amssymb,
 aps,
prb,
]{revtex4-2}

\usepackage[colorlinks = true,
            linkcolor = blue,
            urlcolor  = red,
            citecolor = red,
            anchorcolor = blue]{hyperref}
\usepackage[T1]{fontenc}
\usepackage[utf8]{inputenc}
\usepackage{graphicx}
\usepackage{dcolumn}
\usepackage{color}
\usepackage{amsmath}
\usepackage{appendix}
\usepackage{MnSymbol}
\usepackage{orcidlink}
\usepackage{glossaries}
\newacronym{CDW}{CDW}{charge-density-wave}
\newacronym{RIXS}{RIXS}{resonant inelastic x-ray scattering}
\newacronym{DMRG}{DMRG}{density matrix renormalization group}
\newacronym{BOW}{BOW}{bond-order wave}
\newacronym{DQMC}{DQMC}{determinant quantum Monte Carlo}
\newacronym{QMC}{QMC}{quantum Monte Carlo}
\newacronym{SSH}{SSH}{Su-Schrieffer-Heeger}
\newacronym{oSSH}{oSSH}{optical Su-Schrieffer-Heeger}
\newacronym{2D}{2D}{two-dimensional}
\newacronym{1D}{1D}{one-dimensional}
\newacronym{FS}{FS}{Fermi surface}
\newacronym{eph}{$e$-ph}{electron-phonon}
\newacronym{HMC}{HMC}{hybrid Monte Carlo}
\newacronym{KVB}{KVB}{Kekul{\'e} Valence Bond}
\newacronym{KVBS}{KVBS}{Kekul{\'e} Valence Bond Solid}
\newacronym{SM}{SM}{semi-metal}
\newacronym{QCP}{QCP}{quantum critical point}
\newacronym{FSS}{FSS}{finite-size scaling}
\newacronym{AFM}{AFM}{antiferromagentic}
\newacronym{SC}{SC}{superconducting}

\begin{document}

\preprint{}
\title{Kekul{\'e} valence bond order in the honeycomb lattice optical Su-Schrieffer-Heeger Model and its relevance to Graphene}

\author{Sohan~Malkaruge~Costa\orcidlink{0000-0002-9829-9017}}
\affiliation{Department of Physics and Astronomy, The University of Tennessee, Knoxville, TN 37996, USA}
\affiliation{Institute of Advanced Materials and Manufacturing, The University of Tennessee, Knoxville, TN 37996, USA\looseness=-1} 
\author{Benjamin~Cohen-Stead\orcidlink{0000-0002-7915-6280}}
\affiliation{Department of Physics and Astronomy, The University of Tennessee, Knoxville, TN 37996, USA}
\affiliation{Institute of Advanced Materials and Manufacturing, The University of Tennessee, Knoxville, TN 37996, USA\looseness=-1} 

\author{Steven~Johnston\orcidlink{0000-0002-2343-0113}}
\affiliation{Department of Physics and Astronomy, The University of Tennessee, Knoxville, TN 37996, USA}
\affiliation{Institute for Advanced Materials and Manufacturing, The University of Tennessee, Knoxville, TN 37996, USA\looseness=-1}

\date{\today}

\begin{abstract}
We perform sign-problem-free determinant quantum Monte Carlo simulations of the optical Su-Schrieffer-Heeger (SSH) model on a half-filled honeycomb lattice. In particular, we investigate the model's semi-metal (SM) to Kekul{\'e} Valence Bond Solid (KVBS) phase transition at zero and finite temperatures as a function of phonon energy and interaction strength. Using hybrid Monte Carlo sampling methods we can simulate the model near the adiabatic regime, allowing us to access regions of parameter space relevant to graphene. Our simulations suggest that the SM-KVBS transition is weakly first-order at all temperatures, with graphene situated close to the phase boundary in the SM region of the phase diagram. Our results highlight the important role bond-stretching phonon modes play in the formation of KVBS order in strained graphene-derived systems.
\end{abstract}

\maketitle

\noindent\textit{Introduction} --- Graphene has undergone extensive investigation since it was first isolated~\cite{Novoselov2004Electric, novoselov2005two}. The low-energy physics of undoped monolayer graphene is characterized by a pair of inequivalent gapless Dirac cones located at the edge of the first Brillouin zone~\cite{Castro2009Electronic}. This electronic structure, together with its two-dimensional character, gives rise to a host of exotic emergent properties, including massless Dirac fermion physics, ballistic transport properties~\cite{du2008approaching}, the half-integer quantum Hall effect~\cite{zhang2005experimental}, and relativistic Klein tunneling~\cite{Beenakker2008Colloquium}. 

Graphene's remarkable properties have tremendous potential for applications, which has motivated extensive studies into how they can be manipulated, e.g., through the application of strain~\cite{naumis2023mechanical} or in Moir{\'e} systems~\cite{Cao2018Correlated, Cao2018Unconventional}. 
An important emergent property in this context is the formation of an insulating \gls*{KVBS} phase~\cite{naumis2023mechanical}. This state preserves time-reversal symmetry but gaps the Dirac cones while folding them back to the $\Gamma$ point. It is also linked to chiral symmetry breaking~\cite{Bao2021Experimental, gutierrez2016imaging}, charge fractionalization~\cite{Hou2007Electron} and other electron and phonon related topological properties~\cite{lin2017competing, wu2016topological, blason2022local, Liu2017Pseudospins}. Theoretical and numerical results show that the \gls*{KVBS} state is not only allowed by symmetry~\cite{Frank2011Possible} but can be stabilized by applied isotropic strain~\cite{Sorella2018Correlation}. Experiments have observed \gls*{KVBS} correlations in a variety of settings, including graphene monolayers on silicon oxide~\cite{eom2020direct} or copper~\cite{gutierrez2016imaging} substrates, in lithium or calcium intercalated multi-layers~\cite{sugawara2011fabrication, kanetani2012intercalated, Bao2021Experimental}, and with lithium adatom deposition~\cite{qu2022ubiquitous, cheianov2009hidden}. The \gls*{SM} to insulating \gls*{KVBS} phase transition 
has also been of great fundamental interest. While Landau mean-field theory predicts a first-order transition at the corresponding \gls*{QCP}, quantum fluctuations associated with coupling to gapless Dirac fermion modes can render the quantum phase transition second-order in the chiral XY universality class~\cite{li2017fermion,xu2018kekule,Li2023Emergent}. In contrast, the finite-temperature transition is expected to remain first order~\cite{Liao2019Valence}.

The theoretical studies mentioned above have primarily focused on models dominated by electronic correlations. However, experimental, {\it ab initio}, and mean-field studies have shown that \gls*{KVBS} correlations are often accompanied by periodic lattice distortions and could be driven by \gls*{eph} coupling to graphene's optical in-plane bond-stretching modes~\cite{zhang2022self, classen2014instabilities, otsuka2024kekule}. This interaction results from the modulation of the hopping amplitudes and has been proposed as a potential pairing mechanism in twisted bilayer graphene~\cite{Wu2018Theory, liao2019twisted}. Nevertheless, its role in establishing the \gls*{KVBS} phase has yet to be firmly established as previous \gls*{QMC} studies of \gls*{eph} interactions have focused on Holstein models with high-energy phonons~\cite{Zhang2019charge, Chen2019charge, bradley2023charge}.

In this letter, we investigate the \gls*{SM}-\gls*{KVBS} transition in the \gls*{oSSH} model on a honeycomb lattice, where the in-plane atomic motion modulates the nearest-neighbor hopping amplitude~\cite{MalkarugeCosta2023comparative, TanjaroonLy2023comparative}. By performing sign-problem-free \gls*{DQMC} simulations, we extract the model's ground-state and finite temperature phase diagrams as a function of phonon frequency and \gls*{eph} coupling strength while fully accounting for the quantum nature of the phonons. Our results suggest the \gls*{SM}-\gls*{KVBS} transition is a weakly first-order phase transition down to the \gls*{QCP}. Importantly, we can simulate parameters directly relevant to graphene, allowing us to situate this material close to the \gls*{KVBS} phase boundary. Our results indicate that \gls*{eph} interactions play a crucial role in forming the \gls*{KVBS} phase in strained graphene and help pave the way toward intentional strain engineering of the electronic properties of graphene-derived systems.\\

\noindent\textit{Model} --- Our model's Hamiltonian is $H = H_e+H_\mathrm{ph.}+H_{e-\mathrm{ph.}}$. 
The first term 
\begin{equation*}
    H_e = -t\sum_{{\bf i},\nu,\sigma}\left[\hat{c}^\dagger_{\mathrm{B},{\bf i},\sigma}\hat{c}^{\phantom\dagger}_{\mathrm{A},{\bf i}+{\bf r}_\nu,\sigma}+\rm{h.c.}\right]-
    \mu \sum_{{\bf i},\gamma,\sigma}
    \hat{c}^\dagger_{\gamma,{\bf i},\sigma}\hat{c}^{\phantom\dagger}_{\gamma,{\bf i},\sigma}
\end{equation*}
is the non-interacting electron tight-binding Hamiltonian, where the operator $\hat c_{\gamma,{\bf i},\sigma}^\dagger$, $(\hat c^{\phantom\dagger}_{\gamma,{\bf i},\sigma})$ creates (annihilates) a spin-$\sigma$ electron in orbital $\gamma$ of unit cell ${\bf i}$. The parameter $t$ specifies the nearest-neighbor hopping amplitude, setting the energy scale in the system, and $\mu$ is the chemical potential. Throughout, we consider a half-filled particle-hole symmetric system ($\mu=0$),  where the \gls*{FS} consists of a pair of Dirac points located at $\mathbf{K}_\pm = \tfrac{2\pi}{a}\left(\tfrac{1}{3}, \pm\tfrac{1}{3\sqrt{3}}\right)$.

The second term in $H$ is the non-interacting lattice Hamiltonian 
\begin{equation*}
    H_\mathrm{ph.}=\sum_{{\bf i},\gamma}\left[\frac{1}{2M}\hat{P}_{{\bf i},\gamma}^2+\frac{1}{2}M\Omega^2 \hat{R}_{{\bf i},\gamma}^2\right]. 
\end{equation*}
Here two optical phonon modes are placed on each site to describe the in-plane motion of the ions in the $x$- and $y$-directions, each with frequency $\Omega$ and ion mass $M$. The associated displacement and momentum operators are $\hat{\mathbf{R}}_{{\bf i},\gamma} = \left( \hat{X}_{{\bf i},\gamma}, \hat{Y}_{{\bf i},\gamma} \right)$ and $\hat{\mathbf{P}}_{{\bf i},\gamma} = \left( \hat{P}_{{\bf i},\gamma,X}, \hat P_{{\bf i},\gamma,Y} \right)$, respectively, with $\hat{R}_{{\bf i},\gamma}^2 = |\hat{\mathbf{R}}_{{\bf i},\gamma}|^2$ and $\hat{P}_{{\bf i},\gamma}^2 = |\hat{\mathbf{P}}_{{\bf i},\gamma}|^2$.

\begin{figure}[t]
    \centering
    \includegraphics[width=\columnwidth]{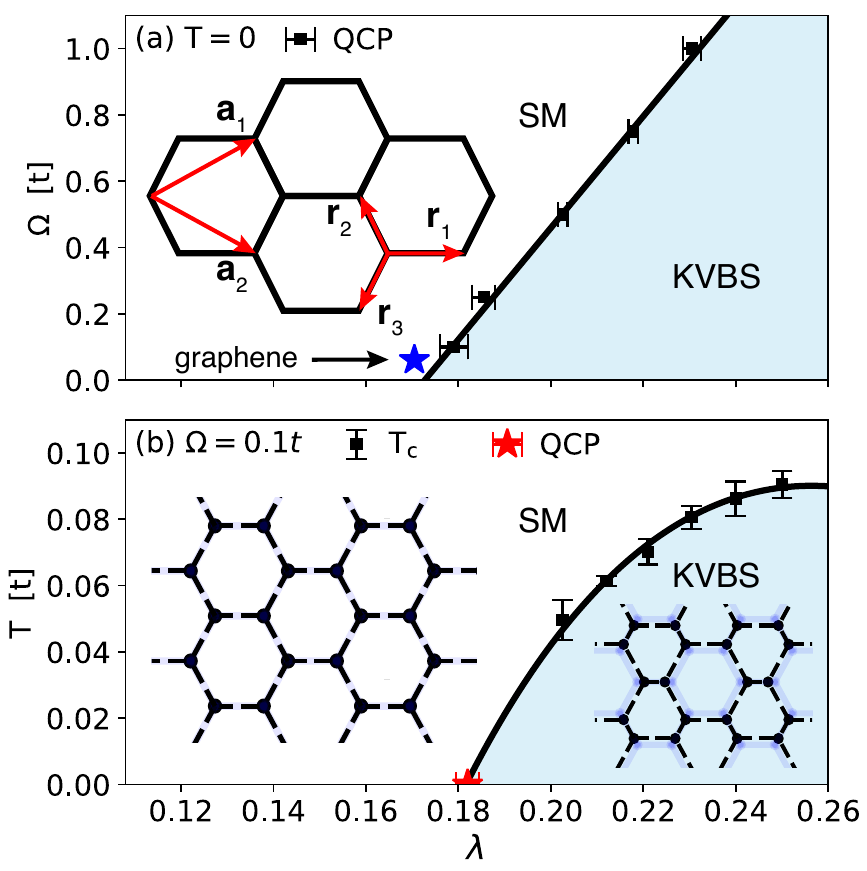}
    \caption{(a) Zero- and (b) finite-temperature phase diagrams for the half-filled \gls*{oSSH} honeycomb model. The left inset in panel a) shows our convention for defining the unit cell. 
    The left and right insets of panel b) panel show an undistorted honeycomb lattice and the \gls*{KVBS} lattice distortion pattern, respectively. The black markers in a) indicate the location of the \gls*{QCP} as a function of \gls*{eph} coupling $\lambda$ and phonon energy $\Omega$. The blue marker is the approximate location of graphene in this zero-temperature phase diagram. The black markers in panel b) are the $T_\mathrm{c}$ values obtained for $\Omega=0.1t$ using the correlation ratio method. The red marker indicates the corresponding \gls*{QCP}. The black curves in both panels are guides for the eye.
    }
    \label{fig:phase}
\end{figure}

Finally, the third term in $H$ describes the \gls*{eph} coupling 
\begin{equation*}
    H_{e-\mathrm{ph.}}=\alpha\sum_{{\bf i},\nu,\sigma} \Delta \hat{R}_{\mathbf{i},\nu} \left[\hat{c}^{\dagger}_{\mathrm{B},{\bf i}+{\bf r}_\nu,\sigma}\hat{c}^{\phantom\dagger}_{\mathrm{A},{\bf i},\sigma}+\rm{h.c.}\right], 
\end{equation*}
where $\Delta \hat{R}_{\mathbf{i},\nu} = \left(\hat{\mathbf{R}}_{{\bf i}+{\bf r}_\nu,\mathrm{B}}-\hat{\mathbf{R}}_{{\bf i},\mathrm{A}}\right) \cdot {\bf r}_\nu / |{\bf r}_\nu|$ and $\mathbf{r}_\nu$ are the three nearest-neighbor vectors shown in Fig.~\ref{fig:phase}(a).
Here, the interaction arises from the linear modulation of the hopping amplitude with the change in bond length, projected onto the equilibrium bond direction $\Delta \hat{R}_{\mathbf{i},\nu}$. The parameter $\alpha$ controls the coupling strength. Throughout, we define the dimensionless ratio $\lambda = \alpha^2/(M\Omega^2 t)$ as a measure of dimensionless coupling, and normalize the mass to $M=1$~\cite{MalkarugeCosta2023comparative}.\\

\noindent\textit{Methods} --- 
We solve our model using \gls*{DQMC} with \gls*{HMC} updates~\cite{Beyl2018revisiting, Batrouni2019langevin, CohenStead2022fast}, as implemented in the \texttt{SmoQyDQMC.jl} package~\cite{cohenstead2024smoqydqmcjl, cohenstead2024smoqydqmcjl_2}. Throughout, we consider clusters with a linear size $L$ with $N = 2 L^2$ orbitals and periodic boundary conditions~\cite{Supplement}.

The \gls*{KVBS} state breaks a $\mathbb{Z}_3$ symmetry and has an electronic order parameter~\cite{Weber2021Valence, xu2018kekule}
\begin{equation}
    \hat{\Psi}_e(\mathbf{K}) = \frac{1}{L^2}\sum_{\mathbf{i},\nu}\left[ e^{-{\rm i}\mathbf{K}\cdot\mathbf{i}} e^{{\rm i} 2\pi\nu/3} \hat{B}_{\mathbf{i},\nu} \right],
\end{equation}
where $\mathbf{K}=\mathbf{K}_+-\mathbf{K}_-$ is the scattering vector between the Dirac points, and $\hat{B}_{{\bf i},\nu} = \sum_{\sigma} (\hat c_{\mathrm{B},{\bf i}+{\bf r}_\nu,\sigma}^\dagger \hat c^{\phantom\dagger}_{\mathrm{A},{\bf i},\sigma} + \text{h.c.})$ is the bond operator. An alternative order parameter can be defined using the lattice displacements
\begin{equation}
    \hat{\Psi}_\text{ph.}(\mathbf{K}) = -\frac{1}{L^2}\sum_{\mathbf{i},\nu}\left[ e^{-{\rm i}\mathbf{K}\cdot\mathbf{i}} e^{{\rm i} 2\pi\nu/3} \Delta \hat{R}_{\mathbf{i},\nu} \right],
\end{equation}
which is sensitive to the distortion pictured in Fig.~\ref{fig:phase}(b). The correspondence between $\hat{\Psi}_e(\mathbf{K})$ and $\hat{\Psi}_\text{ph.}(\mathbf{K})$ occurs because $ \hat{B}_{\mathbf{i},\nu} \sim -\Delta \hat{R}_{\mathbf{i},\nu}$ in the model.

To detect the \gls*{KVBS} state, we measure the bond structure factor $S_\text{vbs}(\mathbf{K}) = L^2 \ \langle \vert \hat{\Psi}_e(\mathbf{K}) \vert^2 \rangle$
and corresponding correlation ratio $R_\text{vbs}(\mathbf{K}) = 1 - S_\text{vbs}(\mathbf{K}+\delta\mathbf{q})/S_\text{vbs}(\mathbf{K})$, 
where $\mathbf{K}+\delta\mathbf{q}$ denotes the nearest momentum points to $\mathbf{K}$ for a given lattice size $L$. In the SM phase, $S_\text{vbs}(\mathbf{k})$ is relatively flat and  $R_\text{vbs}({\bf K}) \rightarrow 0$. Conversely, the bond structure factor will become peaked at ${\bf K}$ as \gls*{KVBS} correlations set in such that $R_\text{vbs}(\mathbf{K}) \rightarrow 1$ below the critical temperature and in the thermodynamic limit.\\

\noindent\textit{Phase Diagram} --- Figure~\ref{fig:phase} shows the zero- and finite-$T$ phase diagrams of the half-filled \gls*{oSSH} model on a honeycomb lattice, which are our main results. In both cases, we see the emergence of a \gls*{KVBS} state from a \gls*{SM} phase, with no indication of additional \gls*{AFM}, charge-density-wave, or superconducting orders~\cite{Supplement}.

Figure~\ref{fig:phase}(a) plots the model's ground-state $(T=0)$ $\lambda$-$\Omega$ phase diagram. The \gls*{KVBS} phase is characterized by both electronic bond correlations and a static periodic distortion of the lattice, pictured in the right inset. These Kekul{\'e} lattice distortions are strongest in the adiabatic limit $(\Omega \rightarrow 0)$ and correspond to the ``Kek-O'' phase characterized by a $\sqrt{3}\times\sqrt{3}$ supercell in which a central hexagon isotropically expands while the six adjacent hexagons contract via an alternating pattern of expanded and contracted bonds. Increasing $\Omega$ enhances quantum fluctuations of the lattice, which disrupt the pattern of Kekul{\'e} lattice distortions and shift the \gls*{SM}-\gls*{KVBS} phase boundary to larger $\lambda$. 

The star in Fig.~\ref{fig:phase}(a) indicates the approximate parameters corresponding to graphene. The relevant phonon modes in graphene are the $A_1$ and $B_1$ modes with energy $\Omega \approx 170~\mathrm{meV} \approx 0.065t$~\cite{Wu2018Theory, Basko2007Effect, Basko2008Interplay}. The lattice distortions associated with the $A_1$ phonon mode are consistent with the \gls*{KVBS} state, and couple to the electrons by modulating the nearest-neighbor hopping amplitude. We estimate the strength of this coupling as  $\alpha \approx 6.09~\text{eV/\AA}$ with a corresponding $\lambda \approx 0.17$~\cite{Supplement}. As expected, monolayer graphene falls in the \gls*{SM} region of the phase diagram but lies very close to the \gls*{SM}-\gls*{KVBS} phase boundary. Thus, small perturbations that increase $\lambda$, as may be expected to accompany the application of an isotropic strain, could drive the emergence of a \gls*{KVBS} state~\cite{Sorella2018Correlation}.

Figure~\ref{fig:phase}(b) shows the model's $\lambda$-$T$ phase diagram for fixed phonon energy $\Omega = 0.1t$, where a phase transition to the \gls*{KVBS} state appears above a critical coupling $\lambda_c$. Previous investigations of this phase transition in purely electronic models indicate that it is weakly first-order at finite temperature but becomes second-order at the \gls*{QCP}, with the quantum phase transition being in the Chiral XY universality class. Our results are consistent with a weakly first-order finite temperature transition in the \gls*{oSSH} model. However, they suggest that the transition remains weakly first-order down to $T=0$ instead of becoming second-order at the \gls*{QCP}.

\begin{figure}[t]
    \centering
    \includegraphics[width=0.75\columnwidth]{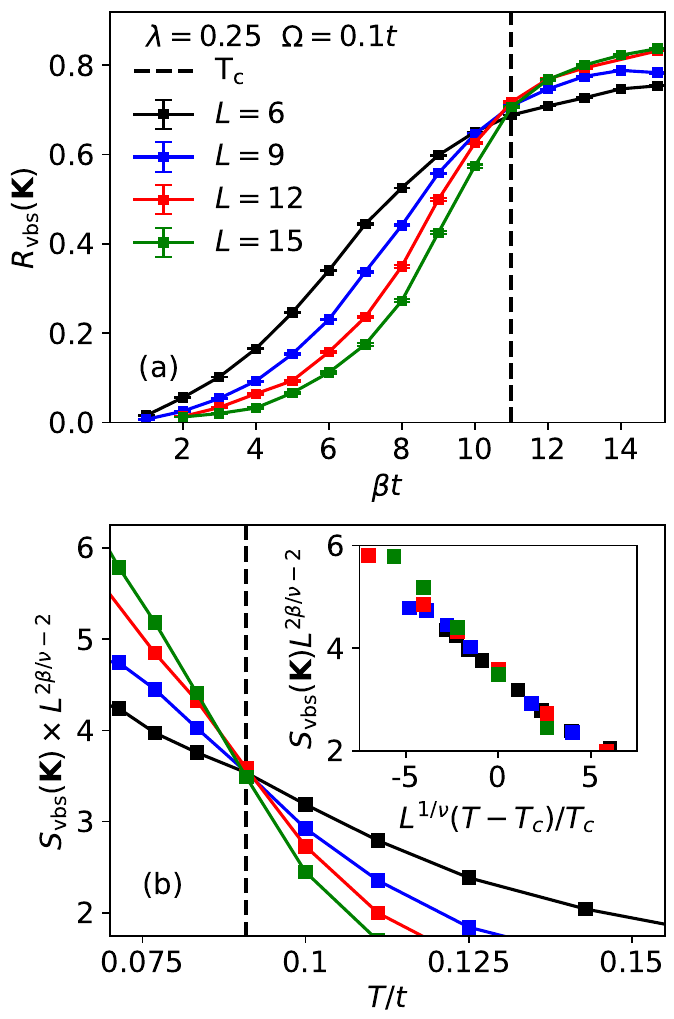}
    \caption{Representative analyses of the finite-temperature SM-KVBS phase transition. a) The temperature evolution of the $R_\text{vbs}(\mathbf{K})$ for different system sizes and $\lambda=0.25$ and $\Omega=0.1t$. The correlation ratio crossing point is approximately $T_\mathrm{c}/t \approx 1/11$, as indicated by the black dashed lines. b) The scaled crossing of $S_\text{vbs}(\mathbf{K})$ versus temperature $T$ using the critical exponents $\beta \approx 0.46$ and $\nu \approx 0.76$, based on the collapse shown in the inset.}
    \label{fig:FSS}
\end{figure}

We first consider the finite temperature transition. Since this transition is expected to be weakly first-order~\cite{xu2018kekule, Zhou2016mott}, we used several approaches to probe its character and determine the critical temperature $T_\mathrm{c}$. Figure~\ref{fig:FSS} presents a \gls*{FSS} analysis of both the the structure factor $S_\text{vbs}(\mathbf{K})$ and correlation ratio $R_\text{vbs}(\mathbf{K})$ for $\Omega = 0.1t$ and $\lambda = 0.25$. Formally, these kinds of \gls*{FSS} analyses are only valid for continuous phase transitions; however, for weakly first-order phase transitions one can obtain a pseudo-scaling behavior for exponents different from the expected universality class but with the correct $T_\mathrm{c}$~\cite{Iino2019detecting}. The crossing point of $R_\text{vbs}(\mathbf{K})$ in Fig.~\ref{fig:FSS}(a) aligns with the one for $S_\text{vbs}(\mathbf{K})$ shown in Fig.~\ref{fig:FSS}(b). The inset in Fig.~\ref{fig:FSS}(b) shows the corresponding collapse, where the critical exponents $\beta \approx 0.46$ and $\nu \approx 0.76$, with an estimated transition temperature of $T_\mathrm{c}/t \approx 1/11$. Given the broken $\mathbb{Z}_3$ symmetry of the \gls*{KVBS} state, these values differ from the three-state Potts model exponents ($\beta = 1/9$ and $\nu = 5/6$) we would expect if the transition were continuous. The transition temperature for other values of $\lambda$ reported in Fig.~\ref{fig:phase}(b) were determined using the crossing point for the correlation ratio given $L=6$, $9$, and $12$.

\begin{figure}[t]
    \centering
    \includegraphics[width=\columnwidth]{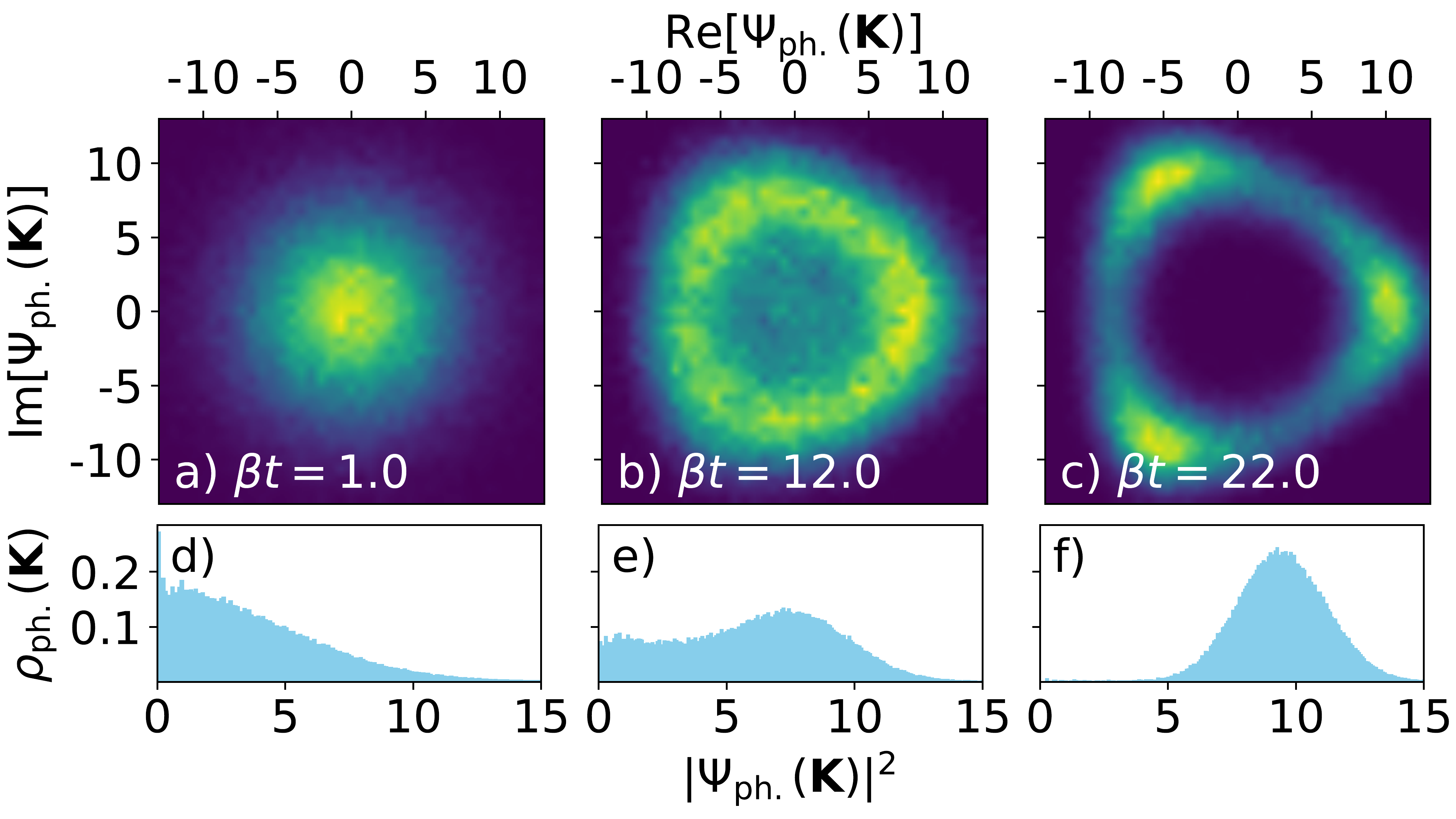}
    \caption{The evolution of the distribution of $\Psi_{\rm lat.}(\mathbf K)$ across the 
    SM to KVBS transition. The results are obtained for $\rm L = 6,\Omega=0.1t,\lambda = 0.22$ simulations. Top row shows results as a 2D histogram for $\rm Im[\Psi_\mathrm{lat.}(\mathbf K)]$ vs $\rm Re[\Psi_\mathrm{lat.}(\mathbf K)]$. The simulation in the SM phase (panel a)  have a central peak. Simulations at the boarder of the SM-KVBS phase transition (panel b), develop a ring-like structure. Simulations in the KVBS phase (panel c) develop 3-lobe structure. The bottom row shows the radial distribution $\rho_\mathrm{lat.}(\mathbf K)$ at  the same $\beta$ values. A clear double peak structure is evident at the boundary between the SM and KVBS phases (panel e), indicating the coexistence of two states and a first-order phase transition. }
    \label{fig:hist}
\end{figure}

To further probe the $\mathbb{Z}_3$ symmetry breaking and provide additional evidence suggesting a weakly first-order finite temperature transition, Fig.~\ref{fig:hist} displays histograms of sampled values of the lattice order parameter $\hat{\Psi}_\text{ph.}(\mathbf{K})$ for $\Omega = 0.1t$ and $\lambda = 0.22$ for an $L=6$ lattice. At $\beta t = 1$ [Fig.~\ref{fig:hist}(a)], well above the transition temperature and absent any symmetry breaking, the sampled $\hat{\Psi}_\text{ph.}(\mathbf{K})$ are symmetrically centered around zero. At $\beta t = 12$ [Fig.~\ref{fig:FSS}(b)], a temperature near the transition, a ring forms around the origin that is not perfectly radially symmetric. There is also significant weight persisting inside the ring, indicative of the fluctuating first-order nature of the \gls*{KVBS} correlations near $T_\mathrm{c}$. At $\beta t = 22$ [Fig.~\ref{fig:FSS}(c)], well below the transition temperature, there is no residual weight near zero and three distinct hot spots have formed on the ring, a result of a broken $\mathbb{Z}_3$ symmetry. Additionally, the location of these three hot spots is consistent with lattice distortions associated with the \gls*{KVBS} state. This behavior is also quite evident in the radial density profiles $\rho_\text{ph.}(\mathbf{K})$ shown in Figs.~\ref{fig:FSS}(d)-(f), which are derived from the histograms shown in panels (a)-(c), respectively.

\begin{figure}[t]
    \centering
    \includegraphics[width=0.95\columnwidth]{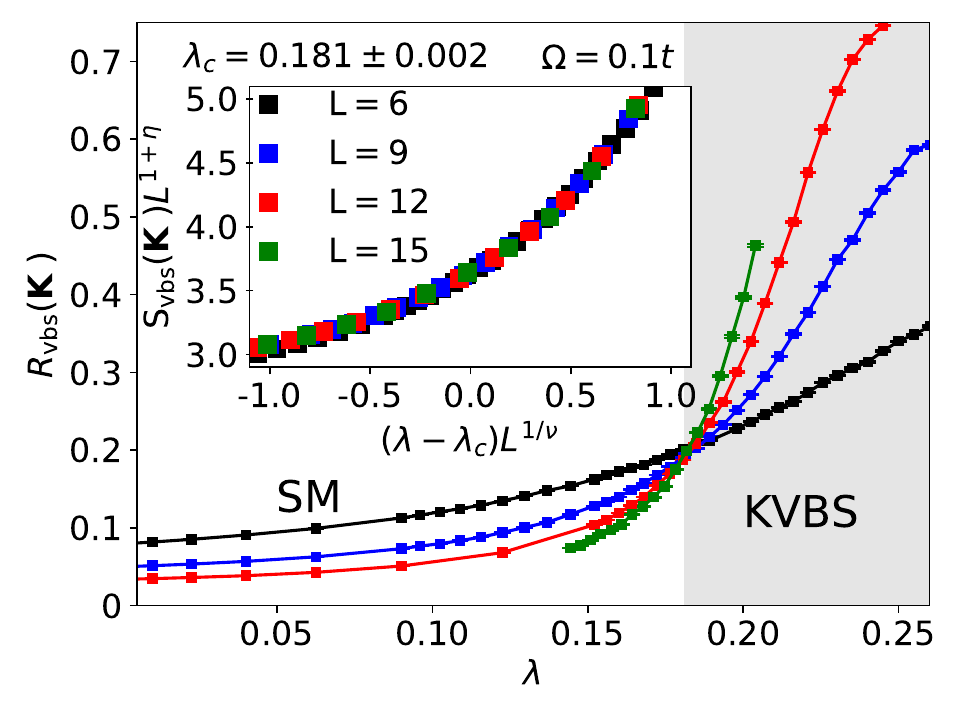}
    \caption{A representative scaling analysis used to obtain  the \gls*{QCP} for \gls*{KVBS} order with $\Omega/t = 0.1$. Here we have considered four lattice sizes with $\beta t=L$, as indicated in the legend. The gray region indicates \gls*{KVBS} order with a critical coupling for the \gls*{SM}-\gls*{KVBS} phase transition of $\lambda_c=0.181\pm0.002$. The inset shows the corresponding finite-size scaling collapse with exponents $\eta \approx -1.16$ and $\nu \approx 0.67$.} 
    \label{fig:QCP}\vspace{-0.25cm}
\end{figure}

To determine the location of the \gls*{QCP} we again perform a \gls*{FSS} using the correlation ratio $R_\text{vbs}(\mathbf{K})$, as shown in Fig.~\ref{fig:QCP}. Holding the phonon frequency $\Omega$ fixed, $\lambda_c$ is given by the crossing point of $R_\text{vbs}(\mathbf{K})$ vs $\lambda$ curves for different lattice sizes $L$. This \gls*{FSS} analysis is based on the emergent Lorentz symmetry in the vicinity of the \gls*{QCP} in which $R_\text{vbs}(\mathbf{K})$ depends on $(\lambda - \lambda_\text{c})/L^\nu$ and $L^z/\beta$, where $z=1$. Therefore, we fix $\beta t = L$ in this analysis  \cite{Assaad2013Pinning,Herbut2009Theory}. Based on the crossing point, we determine that the critical coupling is $\lambda_c = 0.181 \pm 0.002$. The inset then shows the collapse of $S_\text{vbs}(\mathbf{K})$ if we appropriately rescale the axis, treating the critical exponents $\beta$ and $\nu$ as free parameters. We find the optimal collapse with $\eta \approx -1.16$ and $\nu \approx 0.67$, values inconsistent with the Chiral XY universality class~\cite{li2017fermion,xu2018kekule, Liao2019Valence}. Performing a similar collapse of $R_\text{vbs}(\mathbf{K})$ gives a consistent estimate for $\nu$ as well~\cite{Supplement}. These observations suggest that the \gls*{QCP} may remain weakly first-order instead of becoming second-order, as observed in similar studies of purely electronic models. The \gls*{QCP} for other values of $\lambda$ reported in Fig.~\ref{fig:phase}(a) were determined using lattice sizes $L=6$, $9$, and $12$.\\

\noindent{\textit{Discussion}} --- We have presented the ground-state and finite-temperature phase diagrams for the half-filled \gls*{oSSH} model on a honeycomb lattice, investigating the \gls*{SM}-\gls*{KVBS} phase transition. For fixed phonon energy $\Omega$ we identify a critical coupling $\lambda_c$ above which the ground-state transitions from a \gls*{SM} to \gls*{KVBS} state. A finite-temperature phase transition to the \gls*{KVBS} state emerges above $\lambda_c$, with $T_c$ increasing monotonically with $\lambda$. We then estimated the location of graphene in the ground-state phase diagram by assuming the nearest-neighbor electron hopping amplitude couples to the high-energy in-plane $A_1$ bond-stretching phonon mode. The corresponding phonon energy and coupling parameters place graphene in the \gls*{SM} region of the phase diagram, but close to the phase boundary. This result provides new insight into why small perturbations to graphene-derived systems that effectively increase $\lambda$, as is expected to accompany the application of isotropic strain, can drive the system into a \gls*{KVBS} state. This has interesting implications for how strain engineering can be used to tune the electronic properties of graphene-based systems.

Our results also suggest that both the ground-state and finite-temperature \gls*{SM}-\gls*{KVBS} phase transitions are weakly first-order, in agreement with Landau mean-field theory predictions. This result is in contrast with similar \gls*{QMC} investigations of the quantum phase transition in purely electronic models, where coupling to gapless Dirac fermion modes render the \gls*{QCP} second-order, in the Chiral XY universality class. We suspect that our results differ from previous studies as a result retardation effects associated with the \gls*{eph} interaction, which suppresses quantum fluctuations and have been shown to result in \glspl*{QCP} becoming first-order in other situations as well~\cite{Sarkar2023Quantum,Samanta2022Phonon,Chandra2020Quantum}. For instance, a recent investigation similarly found that lattice fluctuations associated with introducing spin-phonon interactions to the honeycomb Heisenberg model drive the deconfined \gls*{QCP} separating the \gls*{AFM} and \gls*{KVBS} ground-states to be first-order~\cite{Weber2021Valence}.

This work focused on half-filling and absent electronic correlations. It would be interesting to introduce a local repulsive Hubbard interaction to study competition between \gls*{SM}, \gls*{KVBS}, and \gls*{AFM} correlations, as the role of \gls*{eph} interactions in this scenario has only been treated at the mean-field level~\cite{otsuka2024kekule}. Another avenue would be to study potential pairing with doping, as prior investigations have found that doping a \gls*{KVBS} state led to superconductivity and pseudogap formation~\cite{Li2023Emergent}.\\

\noindent{\it Acknowledgements} --- 
This work was supported by the U.S.~Department of Energy, Office of Science, Office of Basic Energy Sciences, under Award Number DE-SC0022311. This research used resources of the Oak Ridge Leadership Computing Facility, a DOE Office of Science User Facility supported under Contract No. DE-AC05-00OR22725.

\appendix

\begin{figure}[h]
    \centering
    \includegraphics[width=\columnwidth]{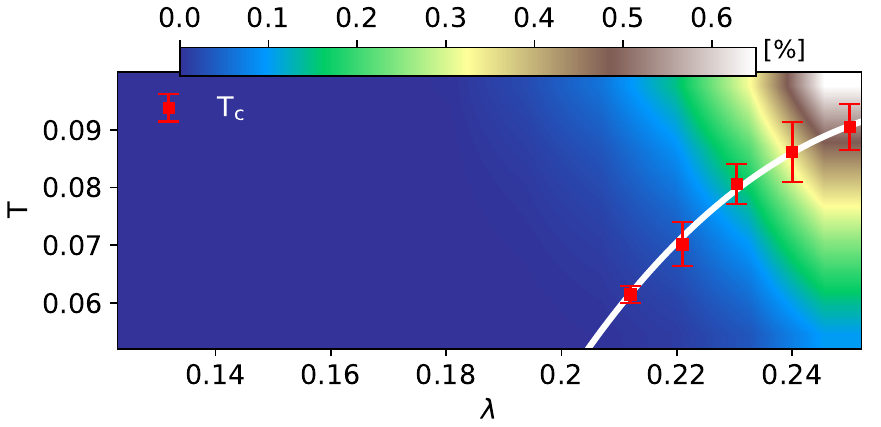}
    \caption{A $\lambda-T$ heat map of the hopping inversion ratio $\eta_{inv}$, measured on an $L=6$ cluster with $\Omega=0.1t$. Markers and the red line indicate the boundary of the SM-KVBS phase transition. The Heat map indicates the Hopping inversion to be lower than 1\% of the total simulation updates. The top right corner shows the bright spot on the heat map, indicating a higher amount of phase inversion occurring in that region.}
    \label{fig:sup_hop_inv}
\end{figure}

\section{DQMC simulations}\label{sec:dqmc_details}
To obtain our results we ran simulations with 12 walkers in parallel, each performing $\sim 5,000$ \gls*{HMC} thermalization updates with $\sim 10,000$ \gls*{HMC} measurement updates. Throughout, we set the imaginary-time discretization to $\Delta\tau = 0.05$. Each \gls*{HMC} update was performed using $N_t \approx 4$ time-steps, with the step-size given by $\Delta t \approx \pi/(2 \Omega N_t)$. We used refer to Ref.~\cite{cohenstead2024smoqydqmcjl} for more information regarding the details of the \gls*{HMC} algorithm.

\section{Parameter estimates for graphene}\label{sec:graphene_approx}
The relevant phonon modes at the $\mathbf{K}_\pm$ Dirac points in graphene are associated with $A_1$ and $B_1$ phonon branches, with corresponding energies of $\hbar\Omega \approx 170~\text{meV}$~\cite{Wu2018Theory, Basko2007Effect, Basko2008Interplay}. The lattice distortions associated with the $A_1$ and $B_1$ phonons are consistent with the Kekul{\'e} lattice distortion pattern observed in our \gls*{DQMC} simulations. To estimate the coupling strength $\alpha$, we consider the functional form for the hopping amplitude 
\begin{equation}\label{eq:tvsd}
    t_\text{eff}(\Delta R) = t e^{-\kappa\frac{\Delta R}{a}} \approx t - \alpha \Delta R + \mathcal{O}(\Delta R^2), 
\end{equation}
which is commonly used in the literature for small changes in the equilibrium bond length $\Delta R$. Here, $a = 1.41~\text{\AA}$ is the lattice parameter in the semi-metal phase, $t = 2.6~\text{eV}$, and $\kappa = 3.3$ \cite{Ribeiro2009Strained}. Thus, the \gls*{eph} coupling strength in the linear approximation is $\alpha = \kappa t/a \approx 6.09~\text{eV/\AA}$. The corresponding dimensionless coupling in our notation is  
\begin{equation}
    \lambda = \frac{\alpha^2}{M\Omega^2 t} = 0.17,
\end{equation}
where $M$ is the mass of carbon and $\hbar\Omega/t = 0.065$.

\begin{figure}[t]
    \centering
    \includegraphics[width=\columnwidth]{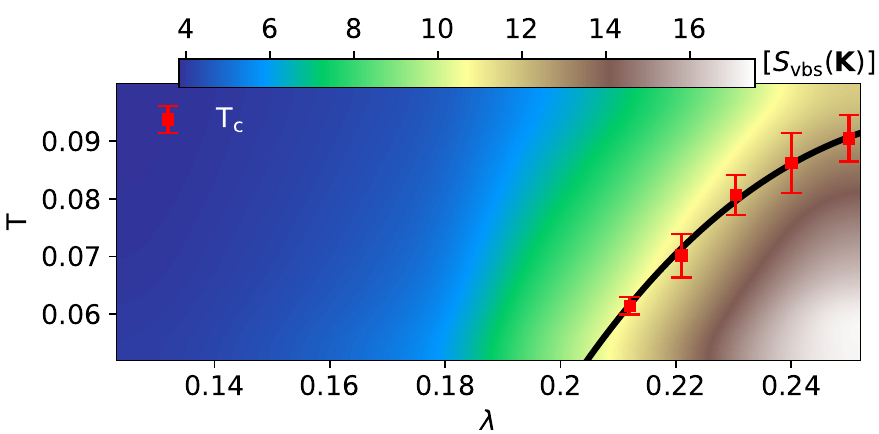}
    \caption{A heat-map of the KVBS structure-factor $S_{\rm vbs}(\mathbf K)$ for $L=6,\Omega=0.1t$, with the black line approximating the location of the SM-KVBS phase boundary. }
    \label{fig:sup_phs_heat}
\end{figure}

\section{Hopping inversion and Phase}
In \gls*{SSH} models, phonon displacements modulate the nearest neighbor hopping amplitude. At linear order in the atomic displacements, the effective hopping is parameterized as in Eq.~\eqref{eq:tvsd} and can change sign for sufficiently large coupling strength $\alpha$ and lattice displacements $\Delta R$. When this occurs, the sign of the effective hopping integral inverts, which induces significant changes in the model's ground and excited state properties~\cite{Banerjee2023groundstate, MalkarugeCosta2023comparative}. It is important to note that such sign changes are not physical but rather a byproduct of the linear approximation for the \gls*{eph} interaction (see Sec.~\ref{sec:graphene_approx}). 

Our implementation of the \gls*{DQMC} algorithm does not place any restrictions on the phonon displacements, so it is possible for $t_\mathrm{eff}$ to be negative. To ensure this does not happen in our model, we measure what percentage of the time this occurs as
\begin{equation}\label{eq:urat}
    \eta_{\rm inv} = 100 \times \langle \Theta(\alpha \hat{R}_{\mathbf{i},\nu}-t) \rangle, 
\end{equation}
where $\Theta$ is the Heaviside function. Fig.~\ref{fig:sup_hop_inv} plots the evolution of this quantity as a function of temperature $T$ and dimensionless coupling $\lambda$ for our simulations performed on an $L = 6$ site cluster. The results 
indicate that a hopping inversion rarely occurs (less than 0.5\%) over the entire range of parameters and that such inversions occur more frequently at high temperatures, where large thermal fluctuations form, or at strong \gls*{eph} coupling. Interestingly, the heat map shows that there is no direct relation between hopping inversion frequency and the formation of the \gls*{KVBS} phase.

\begin{figure*}[t]
    \centering
    \includegraphics[width=0.75\textwidth]{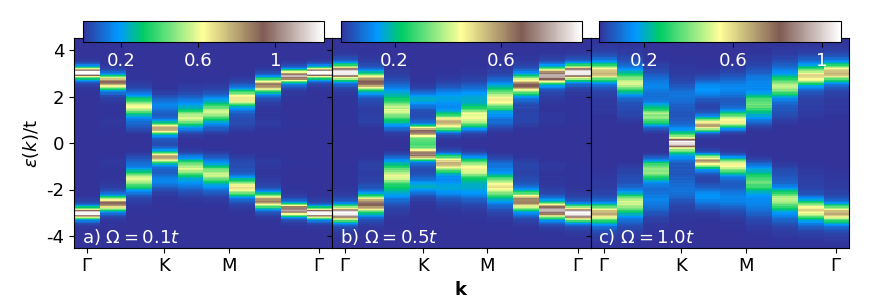}
    \caption{The spectral function $A(\mathbf k,\omega)$ for $L=9$ lattice under $\lambda=0.25$ for different phonon frequency $\Omega$. For panel a) and b) the formation of a KVBS state result in a gap opening at the \gls*{FS}.}
    \label{fig:spect}
\end{figure*}

\section{Additional information on the SM-KVBS phase transition}

The \gls*{KVBS} structure factor may be expressed as
\begin{align}\nonumber
    S_{\rm vbs}(\mathbf K) =&L^2 \ \langle \vert \hat{\Psi}_e(\mathbf{K}) \vert^2 \rangle =\sum_{\nu}S_B^{\nu,\nu}(\mathbf K)\\\label{eq:s3}
    &-\sum_{\nu<\nu'}\mathrm{Re}\left[S_B^{\nu,\nu'}(\mathbf K)\right]\\\nonumber
    &-\sqrt{3}\sum_{\nu<\nu'}(-1)^{\nu+\nu'+1}\mathrm{Im}\left[S_B^{\nu,\nu'}(\mathbf K)\right], 
\end{align}
where
\begin{equation}
    S_B^{\nu,\nu'} = \frac{1}{L^2}\sum_{\mathbf{i},\mathbf{r}} e^{-{\rm i}\mathbf{K}\cdot \mathbf{r}}\langle \hat{B}_{\mathbf{i}+\mathbf{r},\nu} \hat{B}_{\mathbf{i},\nu'}\rangle
\end{equation}
is the equal-time bond-structure factor between bonds $\nu$ and $\nu'$, measured at the scattering momentum $\mathbf K$ connecting the two Dirac points. Fig.~\ref{fig:sup_phs_heat} plots a heat map $S_\mathrm{vbs}(\mathbf K )$ for an $L=6$ lattice with $\Omega=0.1t$. The red line indicates the best fit line we obtain for transition temperature $T_\mathrm{c}$ of the \gls*{SM}-\gls*{KVBS} transition. The heat-map matches well with the estimated $T_\mathrm{c}$ values, with larger $S_{\rm vbs}(\mathbf K)$ values appearing in the \gls*{KVBS} phase.

\section{Results for the electron spectral function}

Figure~\ref{fig:spect} provides supplementary results for the single-particle spectral function $A({\mathbf k},\omega)$ as a function of phonon energy $\Omega$ and a fixed \gls*{eph} coupling $\lambda = 0.25$. For $\Omega = 0.1t$ (Fig.~\ref{fig:spect}a), the system is in the \gls*{KVBS} phase and a gap forms in $A(\mathbf k,\omega)$ at the Dirac point. As the phonon energy is increased, the size of the gap decreases for $\Omega = t/2$ (Fig.~\ref{fig:spect}b) and 
ultimately closes for $\Omega = t$ (Fig.~\ref{fig:spect}c). These results are in line with the theoretical prediction that this particular \gls*{KVBS} order (``Kek-O'' order) induces a gap at the Dirac point. 

\section{Absence of competing charge-density-wave or antiferromagnetic correlations}

\begin{figure}[t]
    \centering
    \includegraphics[width=\columnwidth]{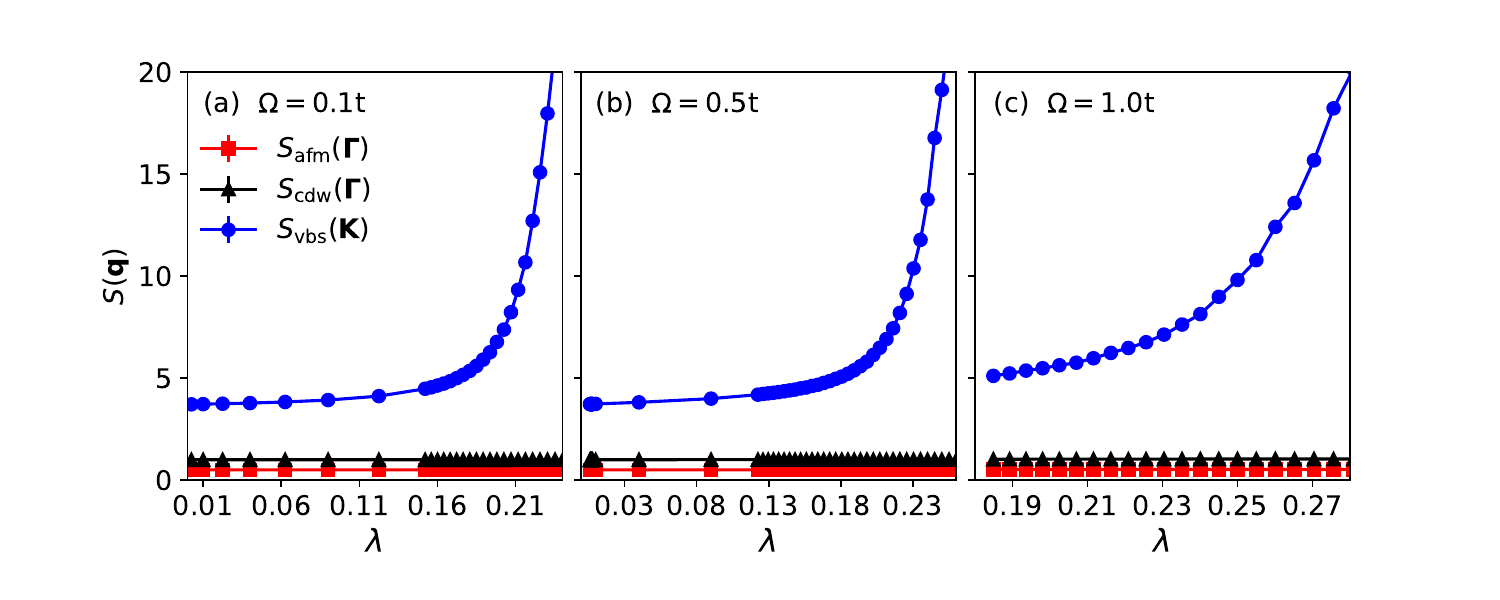}
    \caption{The evolution of the AFM, CDW, and KVBS structure factors as a function of the \gls*{eph} coupling strength $\lambda$ for fixed $T = t/12$. Results are shown for (a) $\Omega = 0.1t$, (b) $\Omega = 0.5t$, and $\Omega = t$. The expected ordering vector for the AFM and CDW correlations is ${\bf \Gamma} = (0,0)$ while the ordering vector ${\bf K}$ for the \gls*{KVBS} correlations is defined in the main text.}
    \label{fig:CDW_AFM_VBS}
\end{figure}

Figure~\ref{fig:CDW_AFM_VBS} shows results for the \gls*{CDW}, \gls*{AFM}, and \gls*{KVBS} structure factors at fixed $T = t/12$ . The measurements show an absence of \gls*{CDW} and \gls*{AFM} phases for the given $\Omega$ range. The \gls*{CDW} and \gls*{AFM} structure factors are given by
\begin{equation}
    S_{\mathrm{cdw}}(\mathbf \Gamma) =\frac{1}{N}\sum_{\mathbf{i},\mathbf{j}}\langle(\hat{n}_{A,\mathbf{i}}-\hat{n}_{B,\mathbf{i}})(\hat{n}_{A,\mathbf{j}}-\hat{n}_{B,\mathbf{j}}) \rangle
\end{equation}
and
\begin{equation}
    S_{\mathrm{afm}}(\mathbf \Gamma) =\frac{1}{N}\sum_{\mathbf{i},\mathbf{j}}\langle(\hat{S}^z_{A,\mathbf{i}}-\hat{S}^z_{B,\mathbf{i}})(\hat{S}^z_{A,\mathbf{j}}-\hat{S}^z_{B,\mathbf{j}}) \rangle,
\end{equation}
respectively, where $\hat{n}_{\gamma,\mathbf{i}}$ and $\hat{S}^z_{\gamma,\mathbf{i}}$ are the total electron number and spin-$z$ operators for orbital $\gamma$ in unit cell $\mathbf{i}$. The definition of $S_{\mathrm{vbs}}(\mathbf{K})$ is provided in the main text. 

\section{Finite Size Scaling Collapse}

\begin{figure}[h]
    \centering
    \includegraphics[width=\columnwidth]{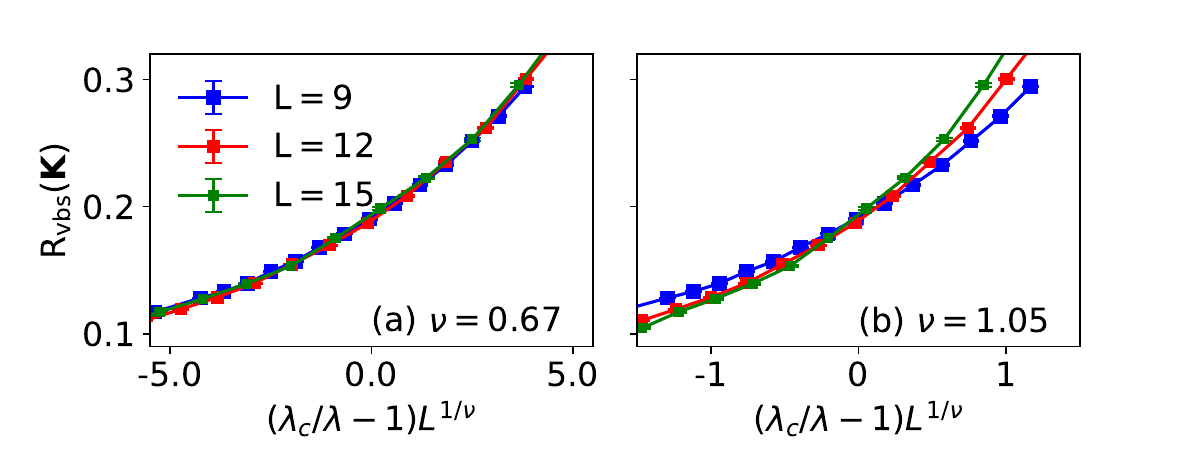}
    \caption{The collapse of $R_\text{vbs}(\mathbf{K})$ for different values of the critical exponent $\nu$. Panel (a) shows that a better collapse can be obtained for $\nu=0.67$ than panel (b), which uses the previously reported value $\nu=1.05$ for the chiral XY universality class.} 
    \label{fig:collapse}
\end{figure}

Figure~\ref{fig:collapse} shows the collapses of $R_\text{vbs}(\mathbf{K})$ using two different values for the critical exponent $\nu$. The left panel is for $\nu = 0.67$, the optimal value obtained based on the $S_\text{vbs}(\mathbf{K})$ collapse reported in the main text. The panel on the right shows the collapse for $\nu = 1.05$, a value previously reported for the chiral XY universality class~\cite{li2017fermion,otsuka2024kekule}. The $\nu = 0.67$  results in a better collapse of the data compared to the $\nu = 1.05$ results.

\bibliography{references.bib}

\end{document}